\documentclass[a4paper,12pt,draft]{article}

\usepackage{amssymb,amsfonts,amsmath,epsfig,float,graphicx}
\usepackage{xcolor}

\usepackage[dvips]{hyperref}
\usepackage[margin=3cm]{geometry}

\title{Non-classical correlations in\\ reducible Quantum Electrodynamics}
\author{Jan Naudts\\
Universiteit Antwerpen\\
\small	\url{Jan.Naudts@uantwerpen.be}\\
\small	\url{https://orcid.org/0000-0002-4646-1190}
}
\date{}

\def\kk{{\bf k}}
\def\Ro{{\mathbb R}}
\newcommand{\be}{\begin{eqnarray}}
\newcommand{\ee}{\end{eqnarray}}
\newcommand{\nnee}{\nonumber\ee}

\newcommand{\upd}{{\rm d}}
\def\kk{{\bf k}}

\def\hc{^\dagger}

\newcommand{\Tr}{\,{\rm Tr}\,}

\def\Io{{\mathbb I}}

\def\Ro{{\mathbb R}}

\def\kk{{\bf k}}

\def\xx{{\bf x}}

\def\kkp{{\bf k'}}

\def\kp{k'}

\def\lu{{\ell}}

\def\ranglec{\rangle^{\mbox{\tiny c}}}
\def\langlec{\,\,\,\strut^{\mbox{\tiny c}}\hskip -1pt\langle}

\def\zetah{\zeta^{\mbox{\tiny H}}}
\def\zetav{\zeta^{\mbox{\tiny V}}}

\def\ah{a_{\mbox{\tiny H}}}
\def\av{a_{\mbox{\tiny V}}}

\def\ahdagger{a\hc_{\mbox{\tiny H}}}
\def\avdagger{a\hc_{\mbox{\tiny V}}}

\def\Nph{N_0 }

\def\Re{\mbox{ Re }}
\def\Im{\mbox{ Im }}

\def\FH{F_{\mbox{\tiny H}}}
\def\FV{F_{\mbox{\tiny V}}}

\begin{document}
\maketitle

\begin{abstract}
The question is discussed whether the momentum of a photon has a quantum uncertainty
or whether it is a classical quantity. The latter assumption is the main characteristic
of reducible Quantum Electrodynamics (rQED). Recent experiments in Quantum Optics may
resolve the question. The non-classical correlation of quantum noise in color-entangled
beams cannot be explained by rQED without modification of the standing explanation.
On the other hand, rQED explains uncertainty of the momentum of a single photon
when it is entangled with a quantum spin residing in its environment. 
The explanation of the historical experiment with equally-polarized pairs of photons,
showing violation of the Bell inequalities,
invokes the argument of collapse of the wave function, also in rQED.
\end{abstract}

\section{Introduction}

In Fock space a superposition of a horizontally polarized photon with wave vector $\kk$
and a vertically polarized photon with wave vector $\kkp$ is described by
\be
\frac{1}{\sqrt 2}\left({\ahdagger}_\kk|0,0\rangle+{\avdagger}_\kkp|0,0\rangle\right).
\label{intro:suppos}
\ee
Here, $|0,0\rangle$ denotes the vacuum state and $\ahdagger$, $\avdagger$
are the creation operators for a horizontally, respectively vertically polarized photon.
Such a superposition with equal vectors, $\kk=\kkp$, 
is for instance needed to describe circularly polarized states.
If the wave vectors are not equal then the superposition describes
a photon whose wave vector and polarization are undetermined.

The existence of color entanglement
has been experimentally established in a convincing manner
\cite{OPKP92,KMWZSS95,VCCMN05,VMFN06,BMPL08,GAMSSL08,CBCVMN09,BdA18}.
The existence of superpositions of the form
\be
\frac{1}{\sqrt 2}\left({\ahdagger}_\kk|0,0\rangle+{\ahdagger}_\kkp|0,0\rangle\right),
\quad \kk\not=\kkp,
\label{intro:suppos2}
\ee
involving a single photon in superposition of two distinct wave vectors,
is required in the analysis of color entanglement as found in \cite{RMD89,SSP02}.
These superpositions describe photon states whose energy and momentum are undetermined.
On the other hand, if (\ref {intro:suppos2}) is not allowed
then the wave vector $\kk$ 
of an idealized plane wave photon is always well-determined and
does not undergo quantum fluctuations.

The present paper questions whether states  of the form (\ref {intro:suppos}) or (\ref {intro:suppos2})
do occur in Nature. If they don't then a modified theory
of Quantum Electrodynamics (QED) is needed.
The original picture behind QED is that Euclidean space is filled with
two-dimen\-sion\-al quantum harmonic oscillators the excitations of which
are photons. Marek Czachor suggested \cite{CM00} that the frequency
of these oscillators should be quantized as well. Together with his
collaborators he developed a non-canonical theory of QED. See
\cite{CM00,CM03,CM04,CN06,CW06,CW09,WC09a,WC09b} and references quoted in these papers.
One of the main results of the theory is that QED after renormalization
is recovered as a limiting case, without the need to fall back on {\em ad hoc}
procedures like renormalization. On the other hand,
an attempt to force the formalism into a mathematically rigorous framework
was not successful \cite{KNunp}.

The formalism of Czachor is still compatible with the existence of
superpositions of the form (\ref {intro:suppos2}). In fact, it emphasizes
the role of these superpositions by assuming that the frequency of the oscillators,
and hence the momentum of the photons, is undergoing quantum fluctuations.
Recently \cite{NJ17,NJ18}, the present author restored the historical
assumption of the frequency being a parameter and added some simplifying assumptions
to the Czachor formalism.
The resulting theory is a reducible version of QED.
It is shortly introduced in the next Section.

The above mentioned removal of superpositions of the form (\ref{intro:suppos}) or (\ref {intro:suppos2})
would of course also effect Quantum Optics. Consider for instance
one-photon states. They are produced experimentally since quite some time
and are used in many experiments \cite {FATBBGZ04,LO05,EFMP11}.
Reducible QED describes realistic one-photon states as superpositions of the form
\be
|\zeta_\kk\rangle&=&c_{0,0}(\kk)|0,0\rangle + c_{1,0}(\kk)|1.0\rangle+c_{0,1}(\kk)|0,1\rangle,
\nnee
with wave vector-dependent coefficients satisfying
\be
|c_{0,0}(\kk)|^2+|c_{1,0}(\kk)|^2+|c_{0,1}(\kk)|^2&=&1,
\nnee
and allows to calculate the corresponding electromagnetic pulse . See Sections \ref{sect:efield}
and \ref {sect:oneph}.

Reducible QED may also lead to a better understanding of the
so-called collapse of the wave function, which is an essential ingredient
in the interpretation of the historical experiments \cite{AGR81,GRA86,HZ86}
on the violation of Bell's inequalities. This point is shortly
discussed at the end of the paper.

The next Section contains a short introduction on rQED.
Sections \ref {sect:efield} and \ref {sect:recon} 
treat the reconstruction of spectral modes following \cite{BCCNFMV13},
but using the axioms of rQED.
Section \ref{sect:polphon} analyzes the historical experiment demonstrating the
entanglement of a photon pair.
Section \ref {sect:covmat} introduces the covariance matrix following \cite{BCCNFVM13}
and shows that with rQED the DGCZ inequalities are not violated by a pair of
identically polarized subbeams. 
Section \ref {sect:env} considers entanglement of a one-photon state
with a spin variable in the environment.
At the end follows a section Discussion and Conclusions.

\section{Reducible QED}

In rQED a quantum electromagnetic field is described by a normalized wave function $\zeta_\kk$
in the Hilbert space $\cal H$ of the two-dimensional quantum harmonic oscillator.
The wave function $\zeta_\kk$ depends on the wave vector $\kk$ which is a non-vanishing
vector in the three-dimensional Euclidean space $\Ro^3$. A relativistic description
is obtained by adding $|\kk|$ as the zeroth component of the 4-momentum.

Let us choose a bazis $|m,n\rangle$ of eigenstates of the two-dimensional quantum harmonic oscillator
and give it the interpretation of describing a state with $m$ horizontally and $n$ vertically
polarized photons. These are photons in the sense of Einstein: interactions
of the electromagnetic field with its
environment occur by creation or annihilation of a photon.
The ground state is $|0,0\rangle$. Excited states are obtained by
the action of the creation operators $\ahdagger$ and $\avdagger$ in the usual manner
\be
\ahdagger|m,n\rangle=\sqrt{m+1}|m+1,n\rangle
\mbox{ and }
\avdagger|m,n\rangle=\sqrt{n+1}|m,n+1\rangle.
\nnee
The annihilation operators $\ah$ and $\av$ are the hermitian conjugates of the
corresponding creation operators.

A superposition of two fields $\zeta_\kk$ and $\eta_\kk$ is of the form 
\be
\lambda\zeta_\kk+\mu \eta_\kk
\mbox{ with } |\lambda|^2+|\mu|^2=1.
\nnee
Wave functions do only combine at equal wave vectors.
For this reason the representation of the Poincar\'e group is reducible.
Quantum averages such as the total energy of the field $\zeta_\kk$
\be
\langle H\rangle_\zeta
&=&
\lu^3\int\upd\kk\,\hbar c|\kk|\,\langle\zeta_\kk|(\ahdagger \ah+\avdagger \av)\zeta_\kk\rangle
\label{rQED:energy}
\ee
decompose into their irreducible components. Note that an arbitrary constant $\lu$
with the dimension of a length has been inserted to make both wave functions and creation
and annihilation operators dimensionless. 
Convergence of the integral in (\ref {rQED:energy}) requires that the wave vectors
$\zeta_\kk$ converge to the vacuum vector $|0,0\rangle$ for large values of $|\kk|$.

A special role is reserved for the coherent electromagnetic fields.
Let $\FH(\kk)$ and $\FV(\kk)$ be two complex functions of the wave vector $\kk$.
Let $|\FH(\kk),\FV(\kk)\ranglec$ denote the corresponding coherent wave function of the
two-dimensional quantum harmonic oscillator. It describes a coherent electromagnetic
field. A short calculation shows that its total energy equals
\be
\langle H\rangle
&=&
\lu^3\int\upd\kk\,\hbar c|\kk|
\langlec\FH(\kk),\FV(\kk)|(\ahdagger \ah+\avdagger \av)|\FH(\kk),\FV(\kk)\ranglec\cr
&=&
\lu^3\int\upd\kk\,\hbar c|\kk|\left[|\FH(\kk)|^2+|\FV(\kk)|^2\right].
\nnee
This gives $\lu^3|\FH(\kk)|^2$ and $\lu^3|\FV(\kk)|^2$ the meaning of the density of
horizontally respectively vertically polarized photons with wave vector $\kk$.

Details of the formalism of rQED are found in the appendices of \cite{NJ18}.

\section{Fluctuations of the electric field operator}
\label{sect:efield}

In rQED the electric field operator at space-time position $x$ is given by
\be
 E_{\alpha,\kk}(x)&=&c|\kk|\frac {\lambda}{2\Nph (\kk)}\varepsilon^{(H)}_\alpha(\kk)
i\left[e^{-ik_\mu x^\mu}\ah-e^{ik_\mu x^\mu}\ahdagger\right]\cr
& &+c|\kk|\frac {\lambda}{2\Nph (\kk)}\varepsilon^{(V)}_\alpha(\kk)
i\left[e^{-ik_\mu x^\mu}\av-e^{ik_\mu x^\mu}\avdagger\right],
\nnee
Here, $\lambda$ is a constant which determines the units in which the electric field is measured.
The polarization vectors  are denoted 
$\varepsilon^{(H)}_\alpha(\kk)$ and $\varepsilon^{(V)}_\alpha(\kk)$.
The relativistic factor $\Nph(\kk)$ is given by
$\Nph(\kk)=\sqrt{(2\pi)^32\lu|\kk|}$, as usually.

The electric field operator can be decomposed into two so-called {\em quadratures}
\be
E_{\alpha,\kk}(x)&=&E^{\mbox{\tiny c}}_{\alpha,\kk}(x)+E^{\mbox{\tiny s}}_{\alpha,\kk}(x)
\nnee
with
\be
E^{\mbox{\tiny c}}_{\alpha,\kk}(x)
&=&c|\kk|\frac {\lambda}{\Nph (\kk)}\cos k_\mu x^\mu
\left\{\varepsilon^{(H)}_\alpha(\kk)\frac 1{2i}(\ahdagger-\ah)
+\varepsilon^{(V)}_\alpha(\kk)\frac 1{2i}(\avdagger-\av)\right\}
\nnee
and
\be
E^{\mbox{\tiny s}}_{\alpha,\kk}(x)
&=&c|\kk|\frac {\lambda}{\Nph (\kk)}\sin k_\mu x^\mu
\left\{\varepsilon^{(H)}_\alpha(\kk)\frac 12(\ah+\ahdagger)
+\varepsilon^{(V)}_\alpha(\kk)\frac 12(\av+\avdagger)\right\}.
\nnee
Fix a field $\zeta(\kk)$. Then the quantum uncertainties are given by
\be
\Delta^2_\zeta E^{\mbox{\tiny c}}_{\kk}(x)
&\equiv&
\sum_\alpha\left\langle\left[E^{\mbox{\tiny c}}_{\alpha,\kk}(x)
-\langle E^{\mbox{\tiny c}}_{\alpha,\kk}(x)\rangle\right]^2\right\rangle_\zeta\cr
&=&
\frac{c^2|\kk|\lambda^2}{(2\pi)^32\lu}\cos^2 k_\mu x^\mu
\left\{
\langle[\Delta\frac{1}{2i}(\ahdagger-\ah)]^2\rangle_\zeta
+\langle[\Delta\frac{1}{2i}(\avdagger-\av)]^2\rangle_\zeta
\right\}
\nnee
and
\be
\Delta^2_\zeta E^{\mbox{\tiny s}}_{\kk}(x)
&\equiv&
\sum_\alpha\left\langle\left[E^{\mbox{\tiny c}}_{\alpha,\kk}(x)
-\langle E^{\mbox{\tiny c}}_{\alpha,\kk}(x)\rangle\right]^2\right\rangle_\zeta\cr
&=&
\frac{c^2|\kk|\lambda^2}{(2\pi)^32\lu}\sin^2 k_\mu x^\mu
\left\{
\langle[\Delta\frac{1}{2}(\ah+\ahdagger)]^2\rangle_\zeta
+\langle[\Delta\frac{1}{2}(\av+\avdagger)]^2\rangle_\zeta
\right\}.
\nnee
In particular, if the wave function $\zeta$ is coherent then 
it follows from
\be
& &\langle\Delta\frac{1}{2i}(\ahdagger-\ah)]^2\rangle_\zeta
=\langle\Delta\frac{1}{2i}(\avdagger-\av)]^2\rangle_\zeta\cr
& &
=\langle[\Delta\frac{1}{2}(\ah+\ahdagger)]^2\rangle_\zeta
=\langle[\Delta\frac{1}{2}(\av+\avdagger)]^2\rangle_\zeta\cr
& &
=\frac 14
\nnee
that
\be
\Delta^2_\zeta E^{\mbox{\tiny c}}_{\kk}(x)
&=&\frac{c^2|\kk|\lambda^2}{(2\pi)^3 4\lu}\cos^2 k_\mu x^\mu
\nnee
and
\be
\Delta^2_\zeta E^{\mbox{\tiny s}}_{\kk}(x)
&=&
\frac{c^2|\kk|\lambda^2}{(2\pi)^3 4\lu}\sin^2 k_\mu x^\mu.
\nnee

The intensity of a light beam is measured with a photo detector.
It produces an electric signal proportional to the intensity.
The quantum uncertainties $\Delta_\zeta E^{\mbox{\tiny c}}_{\kk}(x)$
and $\Delta_\zeta E^{\mbox{\tiny s}}_{\kk}(x)$ of the electric field
result in a contamination of the electric signal with noise. 
In the case of a coherent field this is shot noise and
its intensity is referred to as the {\em quantum noise level}. 
The noise is a direct evidence of the quantum nature of light.
In particular, squeezing of light results in a reduction of noise in
one of the quadratures. Reid \cite{RMD89} proposed to measure
the noise levels of the two quadratures 
in order to demonstrate the Einstein-Podolsky-Rosen (EPR) paradox.
The analysis of noise levels is considered later on in Sections 
\ref {sect:covmat} and \ref {sect:env}.

\section{Reconstruction of spectral modes}
\label{sect:recon}

\def\zetas{\zeta^{\mbox{\tiny s}}}
\def\kkLO{{\kk^{\mbox{\tiny{LO}}}}}
\def\kLO{k^{\mbox{\tiny{LO}}}}

The photon current $I(x)$ measured by a detector at space-time position $x$
is proportional to the expectation $\langle E^{(-)}E^{(+)}\rangle_\zeta$,
where $ E^{(\pm)}$ are the positive and negative frequency parts of the electric field operator.
For a given field $\zeta$ one obtains
\be
I(x)
&\sim&
\lu^3
\int\upd\kk \,
\int\upd\kkp \,
\sum_{\alpha}
\langle\zeta_\kk|E^{(-)}_{\alpha,\kk}(x)E^{(+)}_{\alpha,\kkp}(x)\zeta_{\kkp}\rangle\cr
&=&
\sum_{\alpha}\langle\theta_\alpha(x)|\theta_\alpha(x)\rangle
\nnee
with
\be
|\theta_\alpha(x)\rangle
&=&\lu^{3/2}\int\upd\kk \,E^{(+)}_{\alpha,\kk}(x)|\zeta_\kk\rangle\cr
&=&
\lu^{3/2}\int\upd\kk \,
\frac {c|\kk|\lambda}{2\Nph (\kk)}e^{-ik_\mu x^\mu}
\left[
\varepsilon^{(H)}_\alpha(\kk)\ah|\zeta_\kk\rangle
+\varepsilon^{(V)}_\alpha(\kk)\av|\zeta_\kk\rangle
\right].
\nnee

Consider now a signal field with wave functions $\zetas_\kk$. It is the field which one wants
to investigate by letting it interfere with a coherent field.
The superposition of both fields is described by wave functions
\be
|\zeta_\kk\rangle&=&\sqrt{1-\epsilon^2}|\FH(\kk),\FV(\kk)\ranglec+\epsilon|\zetas_\kk\rangle,
\nnee
where $\FH(\kk)$ and $\FV(\kk)$ are complex functions and $0<\epsilon\ll 1$ is a small constant.
One obtains
\be
|\theta_\alpha(x)\rangle&=&
\lu^{3/2}\int\upd\kk \,
\frac {c|\kk|\lambda}{2\Nph (\kk)}e^{-ik_\mu x^\mu}
\left[
\varepsilon^{(H)}_\alpha(\kk)\FH(\kk)+\varepsilon^{(V)}_\alpha(\kk)\FV(\kk)\right]\,
|\FH(\kk),\FV(\kk)\ranglec\cr
& &
+\epsilon\lu^{3/2}\int\upd\kk \,
\frac {c|\kk|\lambda}{2\Nph (\kk)}e^{-ik_\mu x^\mu}
\left[
\varepsilon^{(H)}_\alpha(\kk)\ah|\zetas_\kk\rangle
+\varepsilon^{(V)}_\alpha(\kk)\av|\zetas_\kk\rangle
\right]\cr
& &
+\mbox{ O }(\epsilon^2).
\nnee
The photon current becomes
\be
I(x)
&\sim&
\lu^3
\int\upd\kk \,\frac {c|\kk|\lambda}{2\Nph (\kk)}e^{ik_\mu x^\mu}
\int\upd\kkp \,\frac {c|\kkp|\lambda}{2\Nph (\kkp)}e^{-i\kp_\mu x^\mu}\sum_{\alpha}\cr
&\times
\bigg\{&
\left[
\varepsilon^{(H)}_\alpha(\kk)\overline{\FH(\kk)}
+\varepsilon^{(V)}_\alpha(\kk)\overline{\FV(\kk)}\right]\,
\left[
\varepsilon^{(H)}_\alpha(\kkp)\FH(\kkp)+\varepsilon^{(V)}_\alpha(\kkp)\FV(\kkp)\right]\,
\cr
& &\qquad\times
\langlec \FH(\kk),\FV(\kk)|\FH(\kkp),\FV(\kkp)\ranglec\cr
& &
+\epsilon
\left[
\varepsilon^{(H)}_\alpha(\kk)\overline{\FH(\kk)}
+\varepsilon^{(V)}_\alpha(\kk)\overline{\FV(\kk)}\right]\,
\cr
& &\qquad\times
\left[
\varepsilon^{(H)}_\alpha(\kkp)\langlec \FH(\kk),\FV(\kk)|\ah\zetas_\kkp\rangle
+\varepsilon^{(V)}_\alpha(\kkp)\langlec \FH(\kk),\FV(\kk)|\av\zetas_\kkp\rangle
\right]\cr
& &
+\epsilon
\left[
\varepsilon^{(H)}_\alpha(\kk)\langle \ah\zetas_\kk|\FH(\kkp),\FV(\kkp)\ranglec
+\varepsilon^{(V)}_\alpha(\kk)\langle\av\zetas_\kk|\FH(\kkp),\FV(\kkp)\ranglec
\right]\,
\cr
& &\qquad\times
\left[
\varepsilon^{(H)}_\alpha(\kkp)\FH(\kkp)+\varepsilon^{(V)}_\alpha(\kkp)\FV(\kkp)\right]\cr
& &
+\mbox{ O }(\epsilon^2)\bigg\}.
\nnee

Relevant information is obtained by spectral analysis of the time dependence of $I(x)$.
Let $A$ denote the set of wave vectors $\kk$ for which $|\FH(\kkp),\FV(\kkp)\ranglec\not=|0,0\rangle$
and $B$ the set of wave vectors $\kk$ for which $\zetas_\kk\not=0$.
Assume that $A$ and $B$ do not overlap and select a frequency $\omega$ such that
$|\kk|-|\kkp|+\omega/c\not=0$ for any pair of wave vectors $\kk,\kkp$ in $A$.
Then the contribution of leading order in $\epsilon$ vanishes and one obtains
\be
\tilde I(\omega)
&\equiv&
\frac{1}{2\pi c}\int\upd x^0\,e^{i\omega x^0/c}I(x)\cr
&\sim&
\epsilon 2\Re
\lu^3
\int_A\upd\kk \,\frac {c|\kk|\lambda}{2\Nph (\kk)}
\int_B\upd\kkp \,\frac {c|\kkp|\lambda}{2\Nph (\kkp)}
e^{-i(\kk-\kkp)\cdot\xx}\delta(\omega+c|\kk|-c|\kkp|)\cr
& &
\times\sum_{\alpha}
\left[
\varepsilon^{(H)}_\alpha(\kk)\overline{\FH(\kk)}
+\varepsilon^{(V)}_\alpha(\kk)\overline{\FV(\kk)}\right]\,
\cr
& &\qquad\times
\left[
\varepsilon^{(H)}_\alpha(\kkp)\langlec \FH(\kk),\FV(\kk)|\ah\zetas_\kkp\rangle
+\varepsilon^{(V)}_\alpha(\kkp)\langlec \FH(\kk),\FV(\kk)|\av\zetas_\kkp\rangle
\right]\cr
& &
+\mbox{O}(\epsilon^2).
\nnee
Experimentally, it is
feasible to measure at once this spectrum and the one with $F(\kk)$ replaced by
$iF(\kk)$ \cite{BCCNFVM13}. In this way also the imaginary part of $\tilde I(\omega)$ is obtained. 
By varying experimental parameters one can obtain values for the functions
$\langlec \FH(\kk),\FV(\kk)|\ah\zetas_\kkp\rangle$ and 
$\langlec \FH(\kk),\FV(\kk)|\av\zetas_\kkp\rangle$.
The coherent states form an overcomplete basis of the Hilbert space of wave functions.
One concludes that, in principle, a reconstruction of the wave vectors
$\ah|\zetas_\kkp\rangle$ and $\av|\zetas_\kkp\rangle$
is feasible. This shows that the results of \cite{BCCNFVM13} translate to the context of rQED.

\section{Entanglement of polarized photon states}
\label {sect:polphon}

Let us convene in the present Section that horizontally polarized photons are independent from
vertically polarized photons. 
Then the wave function $|\zeta\rangle$ of the two-dimensional
harmonic oscillator is said to be separable when it can be written in the form 
$|\zetah\rangle\otimes|\zetav\rangle$.

Assume a two-photon field of the form
\be
|\zeta_\kk\rangle&=&c_{00}|0,0\rangle
+c_{2,0}|2,0\rangle+c_{11}|1,1\rangle+c_{02}|0,2\rangle,
\nnee
with wave vector-dependent complex coefficients $c_{ij}$ satisfying $\sum|c_{ij}|^2=1$.
Assume two sites $x$, $y$ in space-time where it is possible to detect a photon
\cite{fn_deBroglie,dBAS68}.
The measurement at position $x$ destroys a photon of either horizontal
or vertical polarization, using one of the following two annihilation operators
\be
\ah^x&=&\cos(\phi)\ah+\sin(\phi)\av
\quad\mbox{detection of H},\cr
\av^x&=&-\sin(\phi)\ah+\cos(\phi)\av
\quad\mbox{detection of V}.
\nnee
The angle $\phi$ takes into account that the basis of polarization can be rotated.
The wave function $|\zeta_\kk\rangle$ can be written in terms of these rotated operators as
\be
|\zeta_\kk\rangle
&=&
\bigg\{d_{00}
+\frac{1}{\sqrt 2}d_{20}((\ah^x)^\dagger)^2+d_{11}(\ah^x)^\dagger(\av^x)^\dagger
+\frac{1}{\sqrt 2}d_{02}((\av^x)^\dagger)^2
\bigg\}|0,0\rangle,
\nnee
with
\be
d_{00}&=&c_{0,0},\cr
d_{20}&=&c_{20}\cos^2(\phi)+\frac{1}{\sqrt 2}c_{11}\sin(2\phi)+c_{02}\sin^2(\phi),\cr
d_{11}&=&c_{11}\cos(2\phi)-\frac{1}{\sqrt 2}[c_{20}-c_{02}]\sin(2\phi),\cr
d_{02}&=&c_{02}\cos^2(\phi)-\frac{1}{\sqrt 2}c_{11}\sin(2\phi)+c_{20}\sin^2(\phi).
\nnee
If a horizontally polarized photon is taken out then the remaining field is
\be
\zeta_\kk\overset{\mbox{\tiny H}}\rightarrow \left[d_{11}(\av^x)^\dagger+d_{20}(\ah^x)^\dagger+\mbox{ scalar }\right]|0,0\rangle.
\nnee
The arrow $\displaystyle\overset{\mbox{\tiny H}}\rightarrow$ indicates the result of the measurement.
In the original basis the above expression reads
\be
\zeta_\kk
&\overset{\mbox{\tiny H}}\rightarrow&
(-\sin(\phi)d_{11}+\cos(\phi)d_{20})|1,0\rangle\cr
& &
+ (\cos(\phi)d_{11}+\sin(\phi)d_{20})|0,1\rangle+\mbox{ scalar }|0,0\rangle.
\label{spatial:result}
\ee

Without restriction assume that the second measurement occurs in the original basis of the
polarization vectors. Then (\ref {spatial:result}) implies that the probability $P(H,H)$ that the
second measurement returns a horizontal polarization is proportional to
\be
P(H,H)&\sim& |-\sin(\phi)d_{11}+\cos(\phi)d_{2,0}|^2,
\nnee
while the probability $P(H,V)$ of a vertical polarization equals
\be
P(H,V)&\sim& |\cos(\phi)d_{11}+\sin(\phi)d_{2,0}|^2.
\nnee
In the case that $d_{11}=0$, this result coincides with that discussed in \cite{HZ86}.
This case is realized for instance when $c_{11}=0$ and $c_{20}=c_{02}$, which means that the two
photons contributing to $\zeta_\kk$ have the same polarization and both polarizations H and V
have the same weight.

Note that the analysis here, as well as in \cite{HZ86}, relies on the assumption that
the detection of a photon causes a collapse of the wave function which prevents a
subsequent measurement to detect the same photon once again. A more thorough analysis
requires the use of measurement theory and is not elaborated here.

\section{The covariance matrix}
\label{sect:covmat}

Here and in the next Section we focus on a polarized light beam.
To simplify the notations all references to the polarization are omitted.

In the presence of interactions with the environment, the electromagnetic
field is described in rQED by a wave vector-dependent density matrix $\rho_\kk$
instead of a wave function $\zeta_\kk$.
Fix two wave vectors $\kk$ and $\kkp$ and consider measurements at wavelength $\kk$
to be independent from measurements at wavelength $\kkp$.
Any observable $A$ at wavelength $\kk$, respectively $\kkp$,
maps onto an operator $A\otimes\Io$, respectively $\Io\otimes A$.
Because the representation is reducible the density matrix $\rho_{\kk,\kkp}$ of the product space
is separable and can be written as 
\be
\rho_{\kk,\kkp}&=&\sum_np_n\sigma_n\otimes\tau_n,
\label{cov:dm}
\nnee
with density matrices $\sigma_n,\tau_n$ and classic probabilities $p_n\ge 0$, $\sum_np_n=1$.
They satisfy
\be
\rho_\kk=\sum_np_n\sigma_n
\quad\mbox{ and }\quad
\rho_\kkp=\sum_np_n\tau_n.
\nnee
Averages are given by
\be
\langle A\rangle_\kk\equiv\Tr\rho_\kk A&=&\sum_np_n\Tr\sigma_n A,\cr
\langle B\rangle_\kkp\equiv\Tr\rho_\kkp B&=&\sum_np_n\Tr\tau_n B,\cr
\langle A\otimes B\rangle&=&\sum_np_n\left(\Tr\sigma_n A\right)\,\left(\Tr\tau_n B\right).
\nnee

Following \cite{BCCNFVM13} we consider a column vector $X$
with 4 elements given by
\be
X^{\mbox{\tiny T}}&=&(p\otimes\Io\quad q\otimes\Io\quad \Io\otimes p\quad \Io\otimes q).
\nnee
The operators $p$ and $q$ are defined by
\be
q=\frac{1}{\sqrt 2}(a+a^\dagger)
\quad\mbox{ and }\quad
p=\frac{i}{\sqrt 2}(a^\dagger-a)
\nnee
and satisfy the commutation relation $[p,q]=-i$.
A covariance matrix is then defined by
\be
{\Sigma}_{i,j}(\kk,\kkp)
&=&\langle X_iX_j\rangle-\langle X_i\rangle\,\langle X_j\rangle.
\label{cov:covmat}
\ee
One has
\be
[\langle X_iX_j\rangle]_{i,j}
&=&
\left(\begin{array}{lccr}
       \langle p^2\rangle_\kk
       &\langle pq\rangle_\kk
       &\langle p\otimes p\rangle
       &\langle p\otimes q\rangle\\
       \langle qp\rangle_\kk
       &\langle q^2 \rangle_\kk
       &\langle q\otimes p\rangle
       &\langle q\otimes q\rangle\\
       \langle p\otimes p\rangle
       &\langle q\otimes p\rangle
       &\langle p^2 \rangle_\kkp
       &\langle pq\rangle_\kkp\\
       \langle p\otimes q\rangle
       &\langle q\otimes q\rangle
       &\langle qp\rangle_\kkp
       &\langle q^2 \rangle_\kkp
      \end{array}\right)
\nnee
and
\be
\langle X^{\mbox{\tiny T}}\rangle
&=&
(\langle p\rangle_\kk\quad \langle q\rangle_\kk\quad \langle p\rangle_\kkp\quad \langle q\rangle_\kkp).
\nnee

The criterion used in the Literature
to decide whether the two subbeams are entangled
is based on violation of the DGCZ inequality \cite{DGCZ00}.
Note that the density matrix $\rho(\kk,\kkp)$, given by (\ref {cov:dm}),
is separable. Therefore the inequality cannot be violated by this density matrix.

Introduce for instance  \cite{VCCMN05} operators $p_-=(p\otimes\Io-\Io\otimes p)/\sqrt 2$ and 
$q_+=(q\otimes\Io+\Io\otimes q)/\sqrt 2$. They satisfy $[q_+,p_-]=i$.
Then the inequality reads
\be
\Delta^2 p_-+\Delta^2 q_+ \ge 1.
\nnee

\subsection*{The coherent case}
Consider the case that the density matrices $\sigma_n$ and $\tau_n$
are orthogonal projections onto coherent states $|F_n(\kk)+iG_n(\kk)\ranglec$,
respectively $|F_n(\kkp)+iG_n(\kkp)\ranglec$, where $F_n(\kk)$ and $G_n(\kk)$
are real functions of the wave vector $\kk$.
Then one calculates
\be
\langle X\rangle&=&\sqrt 2\sum_n p_n Y_n
\nnee
with
\be
Y_n^{\mbox{\tiny T}}&=&( G_n(\kk)\quad F_n(\kk)\quad G_n(\kkp)\quad F_n(\kkp)).
\nnee
and
\be
[\langle X_iX_j\rangle]_{i,j}
&=&
{\Sigma}^{(0)}+2\sum_np_nY_nY^{\mbox{\tiny T}}_n,
\nnee
with
\be
{\Sigma}^{(0)}&=&
\frac 12
\left(\begin{array}{lccr}
1 &-i &0 &0\\
i &1  &0 &0\\
0 &0  &1 &-i\\
0 &0  &i &1
\end{array}\right).
\nnee
One concludes that in this case the covariance matrix ${\Sigma}(\kk,\kkp)$
equals ${\Sigma}^{(0)}$. In particular, the two subbeams are not correlated.
From
\be
\langle \Delta^2 p_-\rangle&=&\frac 12{\Sigma}_{11}+\frac 12{\Sigma}_{33}-{\Sigma}_{13},\\
\langle \Delta^2 q_+\rangle&=&\frac 12{\Sigma}_{22}+\frac 12{\Sigma}_{44}+{\Sigma}_{24}
\nnee
it follows that in the case of a superposition of coherent fields
one has $\Delta^2 p_-=\Delta^2 q_+=1/2$. Hence, in this case the DGCZ inequality is
actually an equality, as expected.

\section{Entanglement with the environment}
\label{sect:env}

The field of density matrices $\rho_\kk$, introduced in the previous Section,
cannot explain all experimental data. This point is discussed in the final Section.
Let us therefore consider an explicit example of a photon field $\zeta_\kk$ entangled
with a Pauli spin in its environment. The state of the system is described by
\be
\lambda^{\uparrow}\zeta^{\uparrow}_\kk\otimes |\uparrow\rangle
+\lambda^{\downarrow}\zeta^{\downarrow}_\kk\otimes |\downarrow\rangle
\nnee
with $\lambda^{\uparrow}$ and $\lambda^{\downarrow}$ complex numbers satisfying 
$|\lambda^{\uparrow}|^2+|\lambda^{\downarrow}|^2=1$.
The quantum expectation of a field operator $A$ is given by
\be
\langle A\otimes\Io\rangle_\kk
&=&|\lambda^{\uparrow}|^2\langle\zeta^{\uparrow}_\kk|A\zeta^{\uparrow}_\kk\rangle
+|\lambda^{\downarrow}|^2\langle\zeta^{\downarrow}_\kk|A\zeta^{\downarrow}_\kk\rangle.
\nnee

Repeat the construction of the
previous Section. Consider the wave vectors $\kk$ and $\kkp$ as belonging to independent
subbeams. Then the quantum expectation of an observable of the form $A\otimes B\otimes\Io$,
where $A$ refers to the wave vector $\kk$ and $B$ to $\kkp$, is given by
\be
\langle A\otimes B\otimes\Io\rangle_{\kk,\kkp}
&=&
|\lambda^{\uparrow}|^2
\langle\zeta^{\uparrow}_\kk|A\zeta^{\uparrow}_\kk\rangle\,
\langle\zeta^{\uparrow}_\kkp|B\zeta^{\uparrow}_\kkp\rangle
+|\lambda^{\downarrow}|^2
\langle\zeta^{\downarrow}_\kk|A\zeta^{\downarrow}_\kk\rangle\,\langle\zeta^{\downarrow}_\kkp|B\zeta^{\downarrow}_\kkp\rangle.
\nnee

\subsection{Coherent case}

Calculate the covariance matrix ${\Sigma}(\kk,\kkp)$, assuming coherent wave functions
\be
|\zeta^{\uparrow\downarrow}_\kk\rangle
&=&
|F^{\uparrow\downarrow}(\kk)+iG^{\uparrow\downarrow}(\kk)\ranglec,
\nnee
with $F^\uparrow(\kk)$, $F^\downarrow(\kk)$, $G^\uparrow(\kk)$, $G^\downarrow(\kk)$
real functions of the wave vector $\kk$.
One finds
\be
\langle X\rangle=|\lambda^{\uparrow}|^2 \sqrt 2Y^\uparrow+|\lambda^{\downarrow}|^2 \sqrt 2Y^\downarrow
\nnee
with
\be
(Y^\uparrow)^{\mbox{\tiny T}}
&=&
( G^{\uparrow}(\kk)\quad F^{\uparrow}(\kk)\quad G^{\uparrow}(\kkp)\quad F^{\uparrow}(\kkp))
\nnee
and a similar definition for $Y^\downarrow$.
The second order expression is
\be
[\langle X_iX_j\rangle]_{i,j}
&=&
{\Sigma}^{(0)}
+2|\lambda^{\uparrow}|^2 Y^\uparrow(Y^\uparrow)^{\mbox{\tiny T}}
+2|\lambda^{\downarrow}|^2 Y^\downarrow(Y^\downarrow)^{\mbox{\tiny T}}.
\nnee
One obtains
\be
{\Sigma}(\kk,\kkp)
&=&
{\Sigma}^{(0)}\cr
&+&
2|\lambda^\uparrow|^2(1-|\lambda^\uparrow|^2)Y^\uparrow(Y^\uparrow)^{\mbox{\tiny T}}\cr
&+&
2|\lambda^\downarrow|^2(1-|\lambda^\downarrow|^2)Y^\downarrow(Y^\downarrow)^{\mbox{\tiny T}}\cr
&-&
2|\lambda^\uparrow|^2\,|\lambda^\downarrow|^2
\left[Y^\uparrow(Y^\downarrow)^{\mbox{\tiny T}}+Y^\downarrow(Y^\uparrow)^{\mbox{\tiny T}}
\right].
\nnee

Now calculate
\be
\langle \Delta^2 p_-\rangle
&=&
\frac 12
+|\lambda^\uparrow|^2
\left[G^{\uparrow}(\kk)-G^{\uparrow}(\kkp)\right]^2
+|\lambda^\downarrow|^2
\left[G^{\downarrow}(\kk)-G^{\downarrow}(\kkp)\right]^2\cr
&-&
\left(
|\lambda^\uparrow|^2\left[G^{\uparrow}(\kk)-G^{\uparrow}(\kkp)\right]
+|\lambda^\downarrow|^2\left[G^{\downarrow}(\kk)-G^{\downarrow}(\kkp)\right]
\right)^2\cr
&\ge&\frac 12.
\nnee
The inequality follows because the function $f(x)=x^2$ is convex
and $|\lambda^{\uparrow}|^2+|\lambda^{\downarrow}|^2=1$.
Similarly is
\be
\langle \Delta^2 q_+\rangle
&=&
\frac 12
+|\lambda^\uparrow|^2
\left[F^{\uparrow}(\kk)+F^{\uparrow}(\kkp)\right]^2
+|\lambda^\downarrow|^2
\left[F^{\downarrow}(\kk)+F^{\downarrow}(\kkp)\right]^2\cr
&-&
\left(
|\lambda^\uparrow|^2\left[F^{\uparrow}(\kk)+F^{\uparrow}(\kkp)\right]
+|\lambda^\downarrow|^2\left[F^{\downarrow}(\kk)+F^{\downarrow}(\kkp)\right]
\right)^2\cr
&\ge&\frac 12.
\nnee
One concludes that the DGCZ inequality is not violated.
This is not unexpected. See the comments in \cite{ML82}.

\subsection{One-photon states}
\label{sect:oneph}

In the other extreme case the wave function is a one-photon state
entangled with a spin state. 
It is described by
\be
\frac{1}{\sqrt 2}\left[c^{\uparrow}(\kk)|1\rangle+\gamma^{\uparrow}(\kk)|0\rangle\right]
\otimes|\uparrow\rangle
+\frac{1}{\sqrt 2}\left[c^{\downarrow}(\kk)|1\rangle+\gamma^{\downarrow}(\kk)|0\rangle\right]
\otimes |\downarrow\rangle,
\nnee
where $c^{\uparrow}(\kk)$ and $c^{\downarrow}(\kk)$ are complex functions
satisfying $|c^{\uparrow\downarrow}(\kk)|\le 1$,
and
\be
\gamma^{\uparrow}(\kk)&=&\sqrt{1-|c^{\uparrow}(\kk)|^2},
\nnee
with a similar definition of $\gamma^\downarrow(\kk)$.
For the sake of simplicity the two spin states have the same weight.

Let $X$ be as before. 
A tedious calculation yields the DGCZ inequality
\be
& &
\frac 12|c^{\uparrow}(\kk)|^2+\frac 12|c^{\uparrow}(\kk)|^4
+\frac 12|c^{\downarrow}(\kk)|^2+\frac 12|c^{\downarrow}(\kk)|^4\cr
& &
+\frac 12|c^{\uparrow}(\kkp)|^2+\frac 12|c^{\uparrow}(\kkp)|^4
+\frac 12|c^{\downarrow}(\kkp)|^2+\frac 12|c^{\downarrow}(\kkp)|^4\cr
&\ge&
\gamma^{\uparrow}(\kk)\gamma^{\downarrow}(\kk)
\left[
\Re c^{\uparrow}(\kk)\Re c^{\downarrow}(\kk)
+
\Im c^{\uparrow}(\kk)\Im c^{\downarrow}(\kk)
\right]\cr
& &
+\gamma^{\uparrow}(\kkp)\gamma^{\downarrow}(\kkp)
\left[
\Re c^{\uparrow}(\kkp)\Re c^{\downarrow}(\kkp)
+
\Im c^{\uparrow}(\kkp)\Im c^{\downarrow}(\kkp)
\right]\cr
& &
+\gamma^{\uparrow}(\kk)\gamma^{\downarrow}(\kkp)
\left[\Re c^{\uparrow}(\kk)\Re c^{\downarrow}(\kkp)
-\Im c^{\uparrow}(\kk)\Im c^{\downarrow}(\kkp)
\right]\cr
& &
+\gamma^{\downarrow}(\kk)\gamma^{\uparrow}(\kkp)
\left[
\Re c^{\downarrow}(\kk)\Re c^{\uparrow}(\kkp)
-\Im c^{\downarrow}(\kk)\Im c^{\uparrow}(\kkp)
\right].
\nnee
Take for instance 
\be
\Re c^{\uparrow}(\kk)=\Re c^{\downarrow}(\kk)=\Re c^{\uparrow}(\kkp)=\Re c^{\downarrow}(\kkp)=u,\cr
\Im c^{\uparrow}(\kk)=\Im c^{\downarrow}(\kk)=-\Im c^{\uparrow}(\kkp)=-\Im c^{\downarrow}(\kkp)=v.
\nnee
The DGCZ inequality becomes
\be
2u^2+2v^2+2(u^2+v^2)^2
&\ge&
4(1-u^2-v^2)(u^2+v^2),
\nnee
which is equivalent with $u^2+v^2\ge 1/3$.
One concludes that for small intensities the inequality can be violated.

\section{Discussion and Conclusions}

\def\IA{I^{\mbox{\tiny A}}}
\def\kkH{{\kk^{\mbox{\tiny H}}}}

Let $A$ and $B$ denote two non-intersecting regions in the space of wave vectors $\kk$.
They define two subbeams with different 'color'. 
Experimental papers on color entanglement observe a non-classical correlation
between the noise signals which show up when the two subbeams are measured each with
their own photo detector. 
The quantum nature of the correlations is established by demonstrating the violation of a
DGCZ inequality \cite{DGCZ00}. See Section \ref{sect:covmat}. 
This theoretical explanation cannot be reproduced here without modification because in rQED
the density matrices $\rho_\kk$ and $\rho_\kkp$, as given by (\ref {cov:dm}),
are separable right from the start.

At first sight one may conclude that rQED is not compatible with experiment.
An easy way out to conciliate both
is to take into account that entanglement with the environment is
unavoidable. As proved in Section \ref {sect:env}, entanglement with a quantum spin
suffices to allow violation of the DGCZ inequalities under certain circumstances.

There is a more radical solution as well.
Section \ref{sect:polphon} discusses the historical experiment \cite{CHSH69,AGR81} demonstrating
the entanglement of a pair of equally polarized photons. The standard
explanation of this experiment relies on the probabilistic interpretation
of quantum mechanics and on the assumption of a collapse of the wave function.
A similar reasoning is used in Section \ref{sect:polphon}.
The assumption is made that the
measurement of a photon by one detector destroys the photon and
modifies the wave function by the action of the annihilation operator $a$.
A subsequent measurement by the second detector yields a correlated result.
The elaboration of this argument requires the study of measurement theory,
which is not considered here.

\paragraph{Conclusions}
In mainstream QED two photons with distinct wave vectors $\kk\not=\kk'$ are treated as distinct particles.
This is a consequence of the canonical commutation relations, which postulate that
the commutator of creation and annihilation operators is proportional to a Dirac
delta function $\delta^{(3)}(\kk-\kk')$. As a consequence it is meaningful to study the
bi-partite entanglement of photons and to interpret
the observed \cite {KMWZSS95} quantum correlations in terms
of entanglement. In reducible QED the entanglement of 
photons with different wave vectors is not allowed.
Never the less, a bipartite experiment can reveal non-classical correlations.
This is the main result of the present paper.

Some difficulties remain. 
The analysis of the historical two-photon experiment \cite{CHSH69,AGR81} invokes
the argument of the collapse of the wave function. The same explanation works
in the context of rQED. See Section \ref {sect:polphon}.
However, a more in depth analysis requires the use of measurement theory,
which is not considered here. 
The main question of the present paper, whether superpositions of photon states
with unequal wave vectors are physical, and hence whether reducible QED, as presented here,
is an incomplete
description of quantum Electrodynamics, is left open.


\section*{}


\begin{thebibliography}{99}

\bibitem{OPKP92}
Z. Y. Ou, S. F. Pereira, H. J. Kimble and K.C. Peng, {\em
Realization of the Einstein-Podolsky-Rosen paradox for continuous variables,}
Phys. Rev. Lett. {\bf 68}, 3663-3666 (1992).

\bibitem{KMWZSS95}
P. G. Kwiat, K. Mattle, H. Weinfurter, A. Zeilinger, A. V. Sergienko, and Y. Shih, {\em
New High-Intensity Source of Polarization-Entangled Photon Pairs,}
Phys. Rev. Lett. {\bf 75}, 4337--4341 (1995).

\bibitem{VCCMN05}
A. S. Villar, L. S. Cruz, K. N. Cassemiro, M. Martinelli, and P. Nussenzveig, {\em
Generation of Bright Two-Color Continuous Variable Entanglement,}
Phys. Rev. Lett. {\bf 95}, 243603 (2005).

\bibitem{VMFN06}
A. S. Villar, M. Martinelli, C. Fabre, and P. Nussenzveig, {\em
Direct Production of Tripartite Pump-Signal-Idler Entanglement in the
Above-Threshold Optical Parametric Oscillator,} 
Phys. Rev. Lett. {\bf 97}, 140504 (2006).

\bibitem{BMPL08}
V. Boyer, A. M. Marino, R. C. Pooser, P. D. Lett, {\em
Entangled Images from Four-Wave Mixing,}
Science {\bf 321}, 544--547 (2008).

\bibitem{GAMSSL08}
N. B. Grosse, S. Assad, M. Mehmet, R. Schnabel, Th. Symul, and Ping Koy Lam, {\em
Observation of Entanglement between Two Light Beams Spanning an Octave in Optical Frequency,}
Phys. Rev. Lett. {\bf 100}, 243601 (2008).


\bibitem{CBCVMN09}
A. S. Coelho, F. A. S. Barbosa, K. N. Cassemiro, A. S. Villar, M. Martinelli, P. Nussenzveig, {\em
Three-Color Entanglement,}
Science {\bf 326}, 823--826 (2009).


\bibitem{BdA18}
R. Bruzaca de Andrade, {\em
An optical parametric oscillator for a light-atomic media interface,}
PhD Thesis, University of S\~ao Paulo, 2018.


\bibitem{RMD89}
M. D. Reid, {\em
Demonstration of the Einstein-Podolsky-Rosen paradox using 
nondegenerate parametric amplification,}
Phys. Rev. A {\bf 40}, 913 (1989).

\bibitem{SSP02}
Ch. Schori, J. L. S\o rensen, and E. S. Polzik, {\em
Narrow-band frequency tunable light source of continuous quadrature entanglement,}
Phys. Rev. A 66, 033802 (2002).


\bibitem{CM00}
M. Czachor, {\em
Non-canonical quantum optics,}
J. Phys. A{\bf 33}, 8081 (2000).

\bibitem{CM03}
M. Czachor, {\em
States of light via reducible quantization}
Phys. Lett. A {\bf 313}, 380--388 (2003).


\bibitem{CM04}
M. Czachor, {\em
Reducible representations of CAR and CCR with possible applications to field quantization,}
J. Nonlin. Math. Phys. {\bf 11S1}, 78--84 (2004);
arXiv:hep-th/0304144v3.

\bibitem{CN06}
M. Czachor, J. Naudts, {\em
Regularization as quantization in reducible representations of CCR,}
Int. J. Theor. Phys. {\bf 46}, 73 (2007).

\bibitem{CW06}
M. Czachor and M. Wilczewski, {\em
Cavity-QED Tests of Representations of Canonical
Commutation Relations Employed in Field Quantization,}
Int. J. Theor. Phys.  {\bf 46} 1215--1228 (2007).

\bibitem{CW09}
M. Czachor, K. Wrzask, {\em
Automatic regularization by quantization in reducible representations of CCR:
Point-form quantum optics with classical sources,}
Int. J. Theor. Phys. {\bf 48}, 2511--2549 (2009).


\bibitem{WC09a}
M. Wilczewski, M. Czachor, {\em
Theory versus experiment for vacuum Rabi oscillations in lossy cavities,}
Phys. Rev. A {\bf 79}, 033836 (2009).

\bibitem{WC09b}
M. Wilczewski, M. Czachor, {\em
Theory versus experiment for vacuum Rabi oscillations in lossy cavities (II): Direct test of uniqueness of vacuum,}
Phys. Rev. A {\bf 80}, 013802 (2009).

\bibitem {KNunp}
M. Kuna and J. Naudts, unpublished.

\bibitem{NJ17}
J. Naudts, {\em
On the Emergence of the Coulomb Forces in Quantum Electrodynamics,}
Adv. High En. Phys., {\bf 2017}, 7232798 (2017).

\bibitem{NJ18}
J. Naudts, {\em
Emergent Coulomb forces in reducible Quantum Electrodynamics,}
to appear in Found. Sci. (2019).

\bibitem{FATBBGZ04}
S. Fasel, O. Alibart, S. Tanzilli,
P. Baldi, A. Beveratos, N. Gisin and H. Zbinden, {\em
High-quality asynchronous heralded single-photon
source at telecom wavelength,}
New Journal of Physics {\bf 6}, 163 (2004).

\bibitem {LO05}
B. Lounis and M. Orrit, {\em 
Single-photon sources,}
Reports on Progress in Physics, {\bf 68}, 5 (2005).

\bibitem{EFMP11}
M. D. Eisaman, J. Fan, A. Migdall, and S. V. Polyakov, {\em
Single-photon sources and detectors,}
Rev. Sc. Instrum. {\bf 82}, 071101 (2011).

\bibitem{AGR81}
A. Aspect, Ph. Grangier, and G. Roger, {\em
Experimental Tests of Realistic Local Theories via Bell's Theorem,}
Phys. Rev. Lett. {\bf 47}, 460 (1981).

\bibitem{GRA86}
P. Grangier, G. Roger and A. Aspect, {\em
Experimental Evidence for a Photon Anticorrelation Effect on a Beam Splitter: A New Light on Single-Photon Interferences,} Europhysics Lett. {\bf 1}, 173--179 (1986).


\bibitem{HZ86}
M. A. Horne, Anton Zeilinger, {\em
Einstein-Podolsky-Rosen Interferometry,}
Ann. N. Y. Acad. Sc. {\bf 480}, 469--474 (1986).



\bibitem{BCCNFMV13}
F. A. S. Barbosa, A. S. Coelho, K. N. Cassemiro, P. Nussenzveig, 
C. Fabre, M. Martinelli, and A. S. Villar,
{\em
Beyond spectral homodyne detection:  complete quantum measurement of spectral
modes of light,}
Phys. Rev. Lett. {\bf 111}, 200402 (2013).

\bibitem{BCCNFVM13}
F. A. S. Barbosa, A. S. Coelho, K. N. Cassemiro, P. Nussenzveig, 
C. Fabre, A. S. Villar, and M. Martinelli,
{\em
Quantum state reconstruction of spectral field modes:  homodyne and resonator detection schemes, }
Phys. Rev. A{\bf 88}, 052113 (2013).


\bibitem{fn_deBroglie}
de Broglie and Andrade e Silva \cite {dBAS68} argue that the intensity of the electric field
determines the probability of detecting a photon. The strength of the electron-photon
interaction in rQED depends indeed on the intensity of the electric field \cite{NJ18}.

\bibitem{dBAS68}
L. de Broglie and J. Andrade e Silva, {\em
Interpretation of a Recent Experiment on Interference of Photon Beams,}
Phys. Rev. {\bf 172}, 1284--1285 (1968).

\bibitem{DGCZ00}
Lu-Ming Duan, G. Giedke, J. I. Cirac and P. Zoller, {\em
Inseparability criterion for continuous variable systems,}
Phys. Rev. Lett. {\bf 84}, 2722-2725 (2000).


\bibitem{ML82}
L. Mandel, {\em
Tests of quantum mechanics based on interference of photons,}
Phys. Lett. {\bf 89}A, 325--326 (1982).


\bibitem{CHSH69}
J. F. Clauser, M. A. Horne, A. Shimony and R. A. Holt, {\em
Proposed Experiment to Test Local Hidden-Variable Theories,} 
Phys. Rev. Lett. {\bf 23} 880 (1969); Erratum Phys. Rev. Lett. {\bf 24}, 549 (1970).

\end{thebibliography}
\end{document}